\documentclass[a4paper,twoside,10pt]{article}
\usepackage[T1]{fontenc}
\usepackage[latin1]{inputenc}
\usepackage{amsmath, amssymb, latexsym, amsthm, mathrsfs}
\usepackage{graphicx}
\usepackage{ulem}
\usepackage{dsfont}
\input{epsf}
\usepackage{showkeys}
\usepackage{hyperref}

\date{}
\def\XXint#1#2#3{{\setbox0=\hbox{$#1{#2#3}{\int}$}
\vcenter{\hbox{$#2#3$}}\kern-.5\wd0}}

\newcommand{\dt}{dt}

\renewcommand{\epsilon}{\varepsilon}
\renewcommand{\phi}{\varphi}

\theoremstyle{plain}

\theoremstyle{remark}

\title{Variants of the Smith-Wilson method with a view towards applications}
\author{Thomas Viehmann
\thanks{MathInf GmbH, tv@mathinf.eu}
}
\begin{document}

\maketitle

\begin{abstract}
We propose two variants of the Smith-Wilson method for practical application in the insurance industry.
Our first variant relaxes the Smith-Wilson energy and can be used to incorporate less reliable market data with a certain weight rather than disregarding it completely.
This is particularly useful for deriving yield curves in the IFRS~17 accounting regime, where there is a mandate to incorporate all available market data.

A second variant incorporates the requirement to reach the ultimate forward rate at a prescribed term into the problem formulation.
This provides a natural way to fulfil the Solvency~II convergence requirement and is more elegant than the current methodology adapting the term-scale parameter to control convergence.
\end{abstract}

\centerline{AMS Subject Classification: 91G80}

In the context of Solvency~II, the industry-standard yield curve
fitting method of Smith and Wilson~\cite{SW} has become the
preferred method for yield-curve extrapolation beyond what are considered terms
with sufficiently liquid swap (or bond) market.
Since the Smith-Wilson method has been chosen as the interest
rate calibration approach for Solvency~II, its properties have been
extensively discussed. For a detailed description of
the practical application for Solvency~II as well as a discussion of advantages and
disadvantages of the method, we refer to \cite{TechNote}. We briefly review key elements of the method and background in \ref{sec_review}

In the recent IFRS~17 standard for the acconting of insurance contracts \cite{IFRS17}
Implementation Guidance B44 requires: \textit{An entity shall maximise the use of observable inputs
and shall not substitute its own estimates for observable market data except as
described in paragraph 79 of IFRS 13 Fair Value Measurement. Consistent with
IFRS 13, if variables need to be derived (for example, because no observable
market variables exist) they shall be as consistent as possible with observable
market variables.} The usual method of discarding all information beyond the last liquid point can be seen as inconsistant with this guidance, potentially limiting the application of te Smith-Wilson method in the context of IFRS~17.
Thus, we develop and solve a weighted variant of the Smith-Wilson formula in Section~\ref{sec_vsw_weighted} and show an example application to derive a discount curve from swap data in Section~\ref{sec_vsw_application}.

A second extension of the Smith-Wilson formula concerns the Solvency~II specification. There, it has been desired that the forward rates reach the
ultimate forward rate at a prescribed term, denoted \(T_2\). In the
current EIOPA methodology specification, this has approximately been achieved by
the ad hoc method of modifying the smoothness parameter \cite[Section~7.D]{EIOPATD}. In Section~\ref{sec_vsw_convergence} we use the
variational interpretation to derive a variant of the Smith-Wilson
method that explicitly includes reversion to the ultimate forward rate
at \(T_2\) in the problem specification.

\section{A brief review of the Smith-Wilson method}
\label{sec_review}
Smith and Wilson \cite{SW} describe the yield curve in terms of
zero-coupon bond prices given as
\begin{equation}
\label{eq_sw_bond_price}
P(t) := e^{-f_\infty t} + \sum_{k=1}^N \zeta_k W(t, t_k).
\end{equation}
Here, \(t\) is the term, \(t_k\) are the times of cash flows of the
calibration instruments, \(f_\infty\) is the (continuously compounded)
ultimate forward rate, \(zeta_k\) are coefficients to Wilson's kernel
functions \(W\).
The kernel functions themselves are defined for \(t,\tau > 0\) as
\begin{equation}
\label{eq_wilsons_function}
W(t,\tau) := e^{-(t+\tau) f_\infty} \left(\alpha \min\{t,\tau\} -
e^{-\alpha \max\{t,\tau\}} \sinh (\alpha \min\{t,\tau\}) \right).
\end{equation}
When fitting a curve from \(N\) zero-coupon bonds (ZCBs) \(P(t_j)\) with
term \(t_j\), the coefficients
\(\zeta_k\) are found as the solution to the linear system
\begin{equation}
\label{eq_smithwilson_coeff_system}
W \zeta = (P(t_j) - e^{-f_{\infty} t_j})_{j=1}^N
\end{equation}
with the symmetric matrix \(W_{jk} := W(t_j, t_k)\).

The ZCB price function minimises the functional
\[
E_\textrm{SW}(P) := \frac{1}{2 \alpha^3} \int_0^\infty \left|\partial_t^2(e^{f_\infty t} P(t))\right|^2 \dt
      + \frac{1}{2 \alpha} \int_0^\infty \left|\partial_t (e^{f_\infty t} P(t))\right|^2 \dt,
\]
subject to fixing the values \(P(t_k) = P_k\) at \(t_k\) and \(P(0)=1\), see e.g.\ \cite{SW}, also
  referred to in \cite[Section~3.1.7]{SheldonSmith}.

Indeed, the functional is convex and straightforward
calculation shows that the kernel functions are the
fundamental solutions with singularity at \(\tau\) to the distributional Euler-Langrange-Equation
\[
\alpha^{-3} \partial_t^4 (e^{f_\infty t} W(t,\tau)) - \alpha^{-1} \partial_t^2 (e^{f_\infty t}
W(t,\tau)) = \lambda \delta_\tau(t)
\]
with appropriate boundary
conditions at \(t=0\) and limiting behaviour at \(t\rightarrow\infty\)
ensuring that the functional is finite and the price vanishes at infinity.
Here \(\delta_\tau\) is the Dirac-distribution at \(\tau\) and
\(\lambda\) is the Langrange-multiplier needed for imposing the
condition that values of cash flows are met.

Note that outside the singular point, the function
\(f(t) := e^{f_\infty t} W(t,\tau)\) satisfies
\[
\alpha^3 \partial_t^4 f = \alpha \partial_t^2 f
\]
and thus is piece-wise -- with the singular points separating the
pieces -- a linear combination
\[
f(t) = a e^{\alpha t} + b e^{-\alpha t} + c t + d.
\]

\section{A smoothed Smith-Wilson formula balancing smoothness with goodness of fit for less reliable market prices}
\label{sec_vsw_weighted}

Recall that Smith-Wilson method exactly matches the prices of the observed instruments, i.e. it is an interpolation/extrapolation method.
While this is often desired, there are cases when we wish to partially relax this hard constraint. One example is the recent IFRS~17 accounting standard for insurance
contracts quoted above. A cornerstone of IFRS~17 is the desire to maximise the use of market information. For the discount curve, this leads to the question what to do
with prices from markets that are not considered to be fully liquid.

In this section we relax the condition that the calibration instruments' prices
need to be fitted exactly. Instead, we add a (weighted) quadratic
penalty term to the functional. This variant of the Smith-Wilson functional is
\[
E_\textrm{VSW}(P) := E_\textrm{SW}(P)
      + \frac{1}{2} \sum_l w_l \left|\sum_k cf_l(t_k) P(0,t_k) - Pr_l \right|^2.
\]
Naturally, it shares many properties of the original functional
\(E_{SW}\). In particular, the absolutely continuous part of the first
variation is the same and the singular part of the first variation is
again supported in the set of cash flow times \(t_k\).

Indeed, recasting the Smith-Wilson problem as a pure minimisation
problem by defining the functional to be infinite whenever instrument
prices are not matched exactly, this augmented Smith-Wilson functional is
the variational (\(\Gamma\)-) limit of the variant \(E_\textrm{vSW}\).

We could thus allow $w_l$ to be positive infinite, with the convention that
if $w_l = \infty$ the corresponding term in the functional is zero if $\sum_k cf_l(t_k) P(0,t_k) = Pr_l$ and inifite otherwise.

As before, minimizers can be written in terms of the Wilson functions and we are interested in those in the form of equation \eqref{eq_sw_bond_price}.

To solve the minimisation problem involving finite weights, we need to determine the value of the Smith-Wilson functional \(E_\textrm{SW}\).
To this end, we introduce the scalar product
\begin{align*}
\langle P_1,  P_2 \rangle_\textrm{SW}
  &=  \frac{1}{2 \alpha^3} \int_0^\infty \partial_t^2(e^{f_\infty t} P_1(t) \partial_t^2(e^{f_\infty t} P_2(t))  \dt \\
  & \quad
      + \frac{1}{2 \alpha} \int_0^\infty \partial_t(e^{f_\infty t} P_1(t)) \partial_t(e^{f_\infty t} P_2(t)) \dt.
\end{align*}

Observing that products involving the flat interest price curve $P_0(t) = \exp(-t f_\infty)$ vanish, we see that for $P(t)$ as in equation \eqref{eq_sw_bond_price}
\[
E_\textrm{SW}(P(t)) = \sum_k \sum_l \zeta_k \zeta_l \langle W(\,.\,, \tau_k), W(\,.\,, \tau_l) \rangle_{SW} =: \sum_k \sum_l \zeta_k \zeta_l \frac{1}{2} EW_{kl}.
\]

The coefficients \(EW_{kl}\) can be computed numerically or by an elementary but somewhat tedious calculation analytically, which we do in the appendix.

Given instrument cash flows $cf^e_i(t_k)$ and prices $Pr^e_i$ for $i=1,...,N_e$ that we want to fix exactly and cash flows $cf^e_i(t_k)$ and prices $Pr^w_i$ for $i = 1,...,N_w$ that we want to fit with corresponding error weights $w_i$, we
define the residual prices
\[
  \widetilde{Pr}^w_i := \sqrt{w_i} \left( Pr^w_i -\sum_k cf^w_i(t_k) \exp(-f_\infty t_k)\right)
\]
and
\[
  \widetilde{Pr}^e_i := Pr^e_i -\sum_k cf^e_i(t_k) \exp(-f_\infty t_k).
\]
We combine the cash flow sizes and Wilson's function in matrices $C^w$ and $C^e$ with entries
\[
C^w_{il} := \sqrt{w_i} \sum_k cf^w_i(t_k) W(t_k, t_l)
\]
\[
C^e_{il} := \sum_k cf^e_i(t_k) W(t_k, t_l).
\]

We want to solve the quadratic minimisation problem
\begin{align*}
E_\textrm{VSW}(P(t)) &= \frac{1}{2} \zeta^T EW \zeta
                       + \frac{1}{2} \left\| C^w \zeta - \widetilde{Pr}^w \right\|^2 \\
  &= \frac{1}{2} \zeta^T (EW + (C^w)^T C^w) \zeta - (\widetilde{Pr}^w)^T C^w \zeta + \frac{1}{2} \|\widetilde{Pr}^w\|^2
\end{align*}
with the constraints
\[
 C^e \zeta = \widetilde{Pr}^e.
\]
The last term does not depend on the optimisation and can be left out.
This is a quadratic minimisation problem with only inequality constraints.
The solution and a Lagrange multiplier $\lambda$ are given by
\[
\begin{pmatrix} \zeta \\ \lambda \end{pmatrix}
= 
\begin{pmatrix} EW + (C^w)^T C^w & (C^e)^T \\ C^e & 0 \end{pmatrix}^{-1}
\begin{pmatrix} (C^w)^T(\widetilde{Pr}^w) \\ \widetilde{Pr}^e \end{pmatrix}.
\]

This allows us to analytically solve the relaxed variant of the Smith-Wilson problem, i.e. to minimise $E_{VSW}$. We consider a practical application example in the next section.

\section{Application to partially liquid markets}
\label{sec_vsw_application}

Typically bootstrapping uses liquid (in the context of Solvency 2 often as part of \textit{Deep Liquid and Transparent})
data points. Liquidity can be measured in terms of bid-ask spreads, outstanding volume (e.g. of bonds), or trade volume.
Usually we have a \textit{hard} criterion for what is considered liquid, say some characteristic $L$ being at least some threshold $T_L$.

Instead of concerning ourselves with only instruments, with $L \geq T_L$ we then compute a liquidity ratio $R_i := \min\{1, L_i / T_L \} \in [0,1]$ for the $i$th instrument under consideration.
We then match prices exactly for instruments with $R = 1$ and approximately with weight $w_i = -C \ln (1 - R_i)$ for some choice of $C > 0$. Other functions giving an increasing mapping of $(0,1)$ onto $(0, \infty)$ would be equally suitable.

In case of Solvency 2 and the EUR currency, EIOPA has deemed tenors of 1 to 10, 12, 15, and 20 years to be liquid.
Although it is not considered fully liquid we would be interested to consider 30 year swaps with a hypothetical $R_i$ of $0.5$.
\footnote{Although the interest rate swap market - an OTC market - has moved to central clearing, it seems to be hard to obtain publically available information on market liquidity. On June 3rd 2019, LCH showed the following YTD average volume of notional by tenor: $71.9\%$ 0-2 years, $16.1\%$ 2-5 years, $7.1\%$ 5-10 years, $4.3\%$ 10-30 years, $0.6\%$ 30+ years.
          The website \url{https://www.lch.com/services/swapclear/volumes} does not seem to go into the specifics, e.g. on which side of the interval the boundary terms are counted.}

\begin{figure}
  \includegraphics[width=8cm]{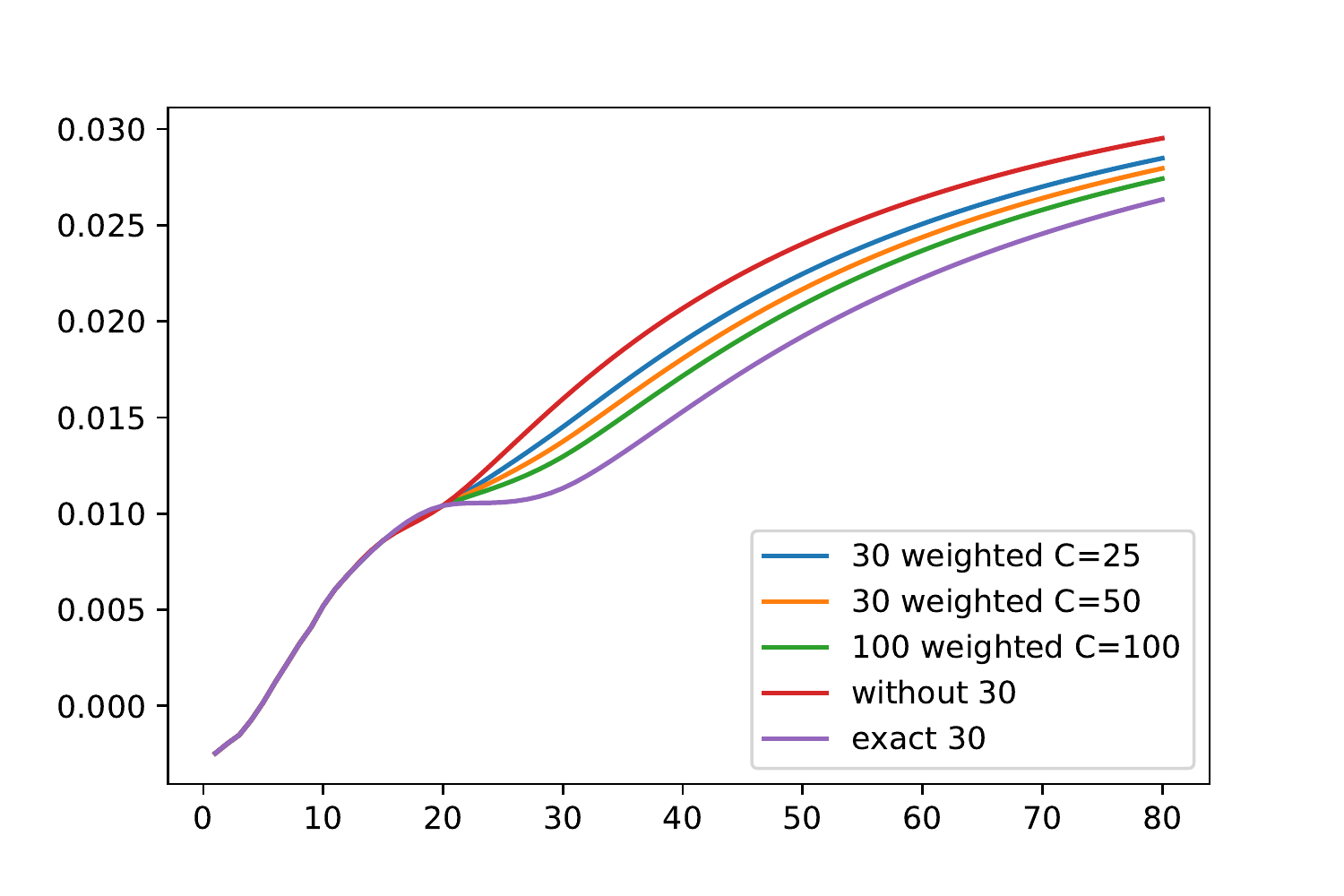}
  \includegraphics[width=8cm]{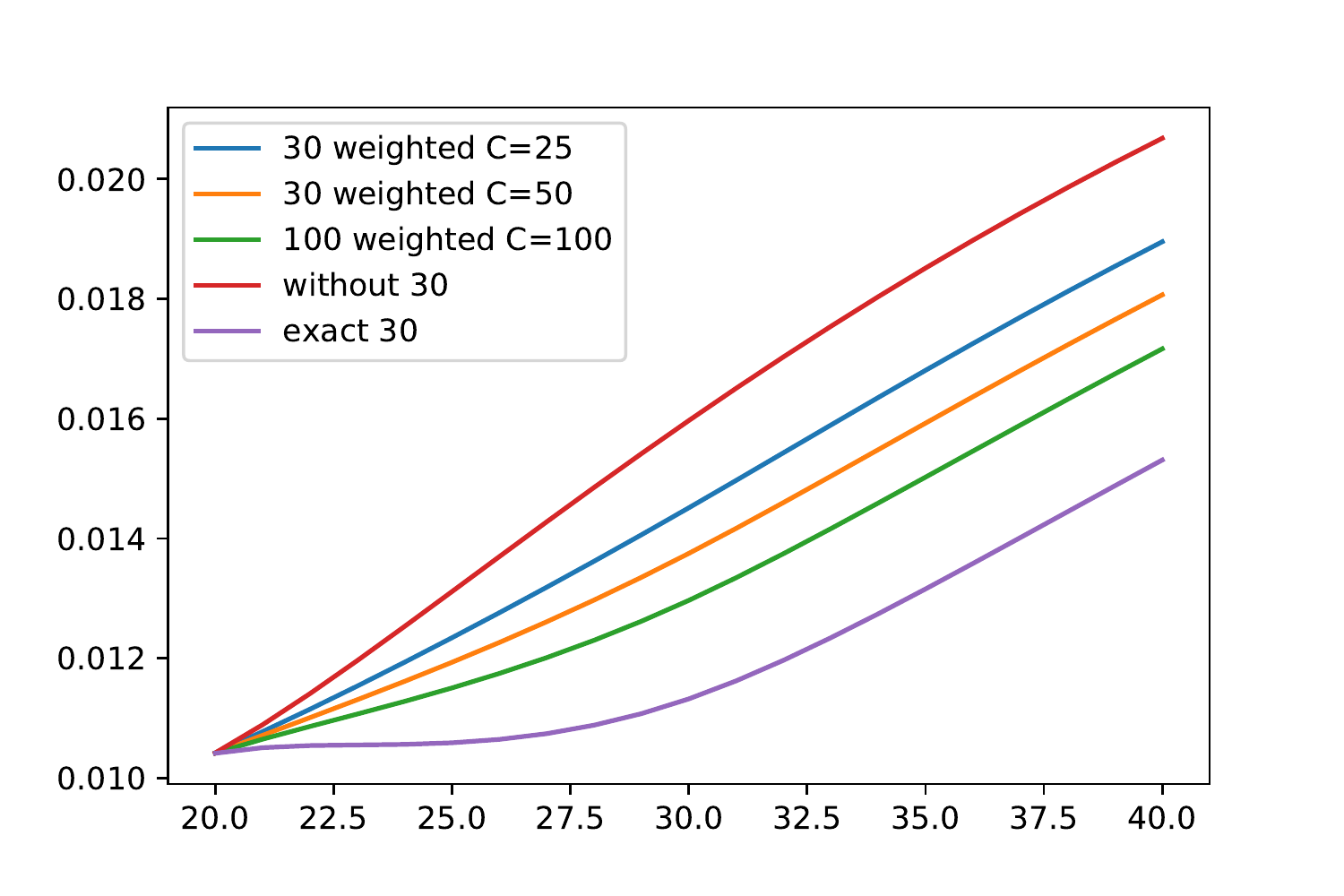}
  \caption{Variations of the spot rate curve given by the Variants of Smith-Wilson. Following EIOPA, the liquid points on the par yield curve are used. The 30 year point is considered partially liquid.
    For comparison, we include the classic Smith-Wilson curve with and without the 30-year point. For all curves we set
    $\alpha=0.1$ and $f_\infty = \ln 1.039$.}
  \label{figure_vsw}
\end{figure}

We illustrate in Figure \ref{figure_vsw} this method for various choices of $C$ using swap data obtained from the Deutsche Bundesbank\footnote{\href{https://www.bundesbank.de/en/statistics/time-series-databases}{Bundesbank Statistical Time Series Databases Zero Coupon Swap Curve}}. As can be seen, the curve is moved towards the 30 year point but does not quite reach it. To understand the parameter, it should be noted that the penalty is on the (unit) price, not the spot yield. Due to taking the 30th root, the spot rate moves much slower than the price.

Our method thus balances the requirement to incorporate market data with the perceived lack of liquidity and thus reliability of the 30 year swap rate.

\section{A variant of the Smith-Wilson method reaching the ultimate
  forward rate after finite time}
\label{sec_vsw_convergence}

In this section we modify the variational problem solved by the
Smith-Wilson interpolating function to incorporate convergence to the
ultimate forward rate at term \(T_2\).

The Smith-Wilson variant function is defined as the minimum of the Smith-Wilson functional
\[
\tilde E_\textrm{SW}(P) := \frac{1}{2 \alpha^3} \int_0^{T_2} \left|\partial_t^2(e^{f_\infty t} P(t))\right|^2 \dt
      + \frac{1}{2 \alpha} \int_0^{T_2} \left|\partial_t (e^{f_\infty t} P(t))\right|^2 \dt,
\]
cut off at $T_2$ among all sufficiently regular Sobolev functions on \((0,T_2)\)
subject to the boundary conditions
\[
P(0)=1, \qquad \frac{\partial_t P}{P} \bigg|_{t=T_2} = -f_{\infty}, \qquad \text{and} \qquad
\partial_t \frac{\partial_t P}{P} \bigg|_{t=T_2} = 0
\]
and the prescribed prices \(P(t_k) = P_k\).

We capture the prescribed prices by introducing the Lagrangian
functional
\[
L_\lambda(P) = \tilde E_{SW}(P) - \sum_{k=1}^N \lambda_k (P(t_k) - P_k).
\]

The boundary conditions at \(t=T_2\) are that the forward rate is
\(f_\infty\) and that the first derivative of the forward rate
vanishes. When extending \(P\) by setting
\(P(t) = e^{-f_\infty  (t-T_2)} P(T_2)\) for \(t>T_2\) we thus have continuity up to the second
derivative of \(P\) and (in general) a jump in the third derivative,
i.e.\ the same regularity as in the original Smith-Wilson function.

We derive the Euler-Lagrange-Equations in the interior by testing with
a smooth function \(\phi\) with compact support in \([0,T_2)\) and \(\phi(0)=0\).
Using partial integration we obtain
\begin{eqnarray*}
\lefteqn{\frac{\partial}{\partial \epsilon} L_\lambda(P+\epsilon \phi)
\bigg|_{\epsilon=0} }
\\ & = &
\alpha^{-3} \int_0^{T_2} \partial_t^2(e^{f_\infty t} P(t))
                      \partial_t^2(e^{f_\infty t} \phi(t)) \dt
      + \alpha^{-1} \int_0^{T_2} \partial_t (e^{f_\infty t} P(t))
      \partial_t (e^{f_\infty t} \phi(t)) \dt
      - \sum_k \lambda_k \phi(t_k)
\\ & = &
\alpha^{-3} \int_0^{T_2} \partial_t^4(e^{f_\infty t} P(t))
                      (e^{f_\infty t} \phi(t)) \dt
      - \alpha^{-3} \partial_t^2 (e^{f_\infty t} P(0)) \partial_t (e^{f_\infty t} \phi(0))
\\ & &
      - \alpha^{-1} \int_0^{T_2} \partial_t^2 (e^{f_\infty t} P(t))
                                          (e^{f_\infty t} \phi(t)) \dt
      - \sum_k \lambda_k \phi(t_k)
\end{eqnarray*}
Testing with \(\phi\) compactly supported in \((0,T_2)\) we thus have the differential equation
\[
\alpha^{-3} \partial_t^4 (e^{f_\infty t} P) - \alpha^{-1} \partial_t^2
(e^{f_\infty t} P) = \sum_{k} \lambda_k \delta_{t_k} \qquad \text{ in } (0,T_2)
\]
in the sense of distributions with \(\delta_{t_k}\) denoting the
Dirac-distribution concentrated at \(t_k\).
Testing with a variation \(phi\) having vanishing value but non-vanishing derivative at
\(t=0\) we obtain the natural fourth boundary condition
\[
 \partial_t^2 (e^{f_\infty t} P(t))\bigg|_{t=0} = 0.
\]
We rewrite the first boundary condition at \(t=T_2\) to have the
homogeneous linear form
\[
\partial_t P \bigg|_{t=T_2} = -f_\infty P(T_2)
\]
Note that the second boundary condition at \(t=T_2\) can be made
linear homogeneous by plugging in the first, i.e.
\[
0 = \partial_t \frac{\partial_t P}{P} \bigg|_{t=T_2} =
\left(\frac{\partial_t^2 P}{P} - \frac{(\partial_t P)^2}{P^2}\right) \bigg|_{t=T_2}
=
\left(\frac{\partial_t^2 P}{P} -  f_\infty^2 \right) \bigg|_{t=T_2} 
\]
implies
\[
\partial_t^2 P\bigg|_{t=T_2} = f_\infty^2 P(T_2) .
\]

The Euler-Lagrange-Equation and the boundary conditions are again
linear. 
Similar to the original Smith-Wilson method we thus can decompose the
minimising function \(P\) as into
\(P_*(t) = e^{-f_\infty t}\) plus a linear combination of kernel
functions with a single Dirac term.

On intervals disjoint from the support of the Dirac measures, the
solution to the fourth-order homogeneous linear differential equation
is a four-dimensional space of functions that can be written as linear combinations
\[
e^{-f_\infty t} (a e^{-\alpha t} + b e^{\alpha t} + c t + d).
\]

With these preparations, we can define the kernel function \(\tilde
W(t,u)\) with singularity at \(u \in (0,T_2)\) as
\[
\tilde W(t,u) =
\begin{cases}
e^{-f_\infty t} (a_0 e^{-\alpha t} + b_0 e^{\alpha t} + c_0 t + d_0) &
\text{for } t \in (0,u) \text{ and} \\
e^{-f_\infty t} (a_1 e^{-\alpha t} + b_1 e^{\alpha t} + c_1 t + d_1) &
\text{for } t \in (u,T_2).
\end{cases}
\]

The boundary condition \(P(0)=1\) translates into \(\tilde W(0,u) =
0\). The other boundary conditions are linear and homogeneous and thus apply to
\(\tilde W(0,u)\) as well.
The boundary conditions at \(t=0\) imply
\begin{eqnarray*}
a_0 + b_0 + d_0 &=& 0, \\ 
a_0 + b_0 &=& 0.      
\end{eqnarray*}
The conditions at \(t=T_2\) yield
\begin{eqnarray*}
 - a_1 \alpha e^{-\alpha T_2} + b_1 \alpha e^{\alpha T_2} + c_1 &=& 0, \\ 
a_1 e^{-\alpha T_2} + b_1 e^{\alpha T_2} &=& 0. 
\end{eqnarray*}
At the singular point \(t=u\) we obtain from identity for the function
and first two derivatives and a jump of height \(\lambda\) in the
third that
\begin{eqnarray*}
  (a_1-a_0) e^{-\alpha u} + (b_1-b_0) e^{\alpha u}+ (c_1-c_0) u +  d_1-d_0 &=& 0, \\ 
- \alpha (a_1-a_0) e^{-\alpha u} + \alpha (b_1-b_0) e^{\alpha u}+ (c_1-c_0) &=& 0, \\ 
  (a_1-a_0) e^{-\alpha u} + (b_1-b_0) e^{\alpha u} &=& 0, \\ 
- (a_1-a_0) e^{-\alpha u} + (b_1-b_0) e^{\alpha u}  &=& \lambda. \\ 
\end{eqnarray*}

Substituting \((1-e^{-2\alpha T_2})\lambda\) for \(\lambda\) and solving for the coefficients, we see
\begin{align*}
a_0 &= \frac{\lambda}{2} (-e^{-2\alpha T_2} e^{\alpha u}-e^{-\alpha u})) &
b_0 &= - \frac{\lambda}{2} (-e^{-2\alpha T_2} e^{\alpha u}-e^{-\alpha u}) \\ 
c_0 &= \lambda \alpha  (1-e^{-2\alpha T_2} - 2 e^{-\alpha T_2} \sinh \alpha u)  &
d_0 &= 0,  \\
a_1 &=   -\lambda \sinh \alpha u, &
b_1 &=   \lambda e^{- 2 \alpha T_2} \sinh \alpha u, \\
c_1 &= - \lambda  e^{- \alpha T_2} \sinh  \alpha u,  &
d_1 &= (1-e^{-2\alpha T_2})\lambda \alpha u.\\
\end{align*}

The function is proportional to \(\lambda\), choosing any \(\lambda = \lambda(u) \neq 0\) will result in the same extrapolation.
Note that the function \(\tilde W\) is not symmetric in the two parameters \(t\) and \(u\) due to the asymmetry of the boundary conditions.

With \(\tilde W(t,u)\) defined, we can now solve equation \eqref{eq_smithwilson_coeff_system} to obtain coefficients and 
use \eqref{eq_sw_bond_price} with \(W\) replaced by \(\tilde W\) to extrapolate the yield curve such that the ultimate forward rate is reached at \(T_2\).
The extension to the calibration to coupon bonds or general series of cash flows is also fully parallel to that with the original Smith-Wilson method, see e.g.\ \cite{TechNote}.

\section{Conclusion}

We present two variants of the Smith-Wilson method of particular practical use enabled by appreciation of the variational nature of the Smith-Wilson method.

The first allows to incorporate less liquid and thus not completely reliable market data. This is a desirable property in the construction of discount curves for IFRS~17.

The second explicitly addresses the desire to reach the ultimate forward rate after finite time, which, in Solvency~II is achieved by a rather unnatural adaptation of the smoothness parameter $\alpha$.

\section*{Acknowledgements}

The author thanks Barbara Blum for helpful comments. All errors are his own.

\newpage

\section*{Appendix: Coefficients for the Smith-Wilson energy}

Here we compute the coefficients for the Smith-Wilson energy needed in Section~\ref{sec_vsw_weighted}.
We consider two indices $k$, $l$ corresponding to cash flow times $\tau_k$ and $\tau_l$.
\begin{align*}
EW_{kl} := 2 \langle W(\,.\, , \tau_k),  W(\,.\, , \tau_l) \rangle_\textrm{SW}
  &=  \frac{1}{\alpha^3} \int_0^\infty \partial_t^2(e^{f_\infty t} W(t, , \tau_k)) \partial_t^2(e^{f_\infty t} W(t , \tau_l))  \dt \\
  & \quad
      + \frac{1}{\alpha} \int_0^\infty \partial_t(e^{f_\infty t} W(t , \tau_k)) \partial_t(e^{f_\infty t} W(t , \tau_l)) \dt.
\end{align*}

Recalling definition of the Wilson function $W$ from above, we take the derivative of
\begin{align*}
e^{t f_\infty} W(t,\tau) &= e^{-\tau f_\infty} \left(\alpha \min\{t,\tau\} -
e^{-\alpha \max\{t,\tau\}} \sinh (\alpha \min\{t,\tau\}) \right)
\end{align*}
to get
\begin{align*}
  \partial_t (e^{t f_\infty} W(t,\tau))
  & = \alpha e^{-\tau f_\infty}
    \begin{cases}
      1 -  e^{-\alpha \tau} \cosh(\alpha t) & \text{ if } t < \tau, \\
      e^{-\alpha t} \sinh (\alpha \tau)  & \text{ if } t > \tau,
  \end{cases}
\end{align*}
and
\begin{align*}
  \partial_t^2 (e^{t f_\infty} W(t,\tau))
  & = - \alpha^2 e^{-\tau f_\infty} e ^{-\alpha \max \{t, \tau\} } \sinh (\alpha \min\{t, \tau\}).
\end{align*}

Without loss of generality, $\tau_k < \tau_l$. Decomposing the integration domain $(0, \infty)$ into open intervals $I_1 = (0, \tau_k)$, $I_2 = (\tau_k, \tau_l)$ and $I_3 = (\tau_l, \infty)$ we have
\begin{align*}
  & \quad \int_0^\infty \partial_t^2(e^{f_\infty t} W(t , \tau_k)) \partial_t^2(e^{f_\infty t} W(t , \tau_l))  \dt \\
  &= \frac{1}{4} \alpha^4 e^{-(\tau_k+\tau_l) f_\infty} \Bigg( e^{- \alpha (\tau_k + \tau_l)} \int_0^{\tau_k}   e^{2 \alpha t} + e^{-2 \alpha t}-2 \, dt \\
  & \qquad \qquad\qquad\qquad  + \int_{\tau_k}^{\tau_l}  e^{\alpha (\tau_k - \tau_l)} -  e^{\alpha (- \tau_k - \tau_l)} + (e^{-\alpha (\tau_k + \tau_l)}  -  e^{\alpha (\tau_k-\tau_l)}) e^{-2\alpha t} dt \\
  & \qquad \qquad\qquad\qquad   +  4\sinh(\alpha \tau_k) \sinh(\alpha \tau_l) \int_{\tau_l}^\infty e^{- 2\alpha t}  dt \Bigg) \\
  &= \frac{1}{4} \alpha^4 e^{-(\tau_k+\tau_l) f_\infty} \Bigg( e^{- \alpha (\tau_k + \tau_l)}  (\frac{1}{2\alpha} (e^{2 \alpha t}- e^{- 2 \alpha t}) - 2t)\Bigg|_{t=0}^{\tau_k} \\
  & \qquad \qquad\qquad\qquad  +  (e^{\alpha (\tau_k - \tau_l)} -  e^{\alpha (- \tau_k - \tau_l)}) t -\frac{1}{2\alpha} (e^{-\alpha (\tau_k + \tau_l)}  -  e^{\alpha (\tau_k-\tau_l)} ) e^{-2\alpha t} \Bigg|_{t=\tau_k}^{\tau_l}  \\
  & \qquad \qquad\qquad\qquad   -  4\sinh(\alpha \tau_k) \sinh(\alpha \tau_l) \frac{1}{2\alpha}  e^{-2\alpha t} \Bigg|_{t=\tau_l}^\infty  \Bigg) \\
\end{align*}
\begin{align*}
  &= \frac{1}{4} \alpha^4 e^{-(\tau_k+\tau_l) f_\infty} \Bigg( e^{- \alpha (\tau_k + \tau_l)}  (\frac{1}{2\alpha} (e^{2 \alpha \tau_k}- e^{- 2 \alpha \tau_k}) - 2\tau_k) \\
  & \qquad \qquad\qquad\qquad  +  (e^{\alpha (\tau_k - \tau_l)} -  e^{\alpha (- \tau_k - \tau_l)}) (\tau_l - \tau_k) -\frac{1}{2\alpha} (e^{-\alpha (\tau_k + \tau_l)}  -  e^{\alpha (\tau_k-\tau_l)} ) (e^{-2\alpha \tau_l}-e^{-2\alpha \tau_k})   \\
  & \qquad \qquad\qquad\qquad  +  \sinh(\alpha \tau_k) \sinh(\alpha \tau_l) \frac{1}{2\alpha}  e^{-2\alpha \tau_l} \Bigg) \\
  &= \frac{1}{4} \alpha^3 e^{-(\tau_k+\tau_l) f_\infty} \Bigg( e^{- \alpha (\tau_k + \tau_l)}  (\sinh(2 \alpha \tau_k) - 2\tau_k\alpha) \\
  & \qquad \qquad\qquad\qquad  + e^{-\alpha \tau_l} 2 \sinh(\alpha \tau_k) \alpha (\tau_l - \tau_k) + \sinh(\alpha \tau_k) e^{-\alpha \tau_l} (e^{-2\alpha \tau_l}-e^{-2\alpha \tau_k})   \\
  & \qquad \qquad\qquad\qquad  +  4\sinh(\alpha \tau_k) \sinh(\alpha \tau_l) \frac{1}{2}  e^{-2\alpha \tau_l} \Bigg)
\end{align*}
for the second derivative and
\begin{align*}
  & \quad \int_0^\infty \partial_t (e^{f_\infty t} W(t , \tau_k)) \partial_t(e^{f_\infty t} W(t , \tau_l))  \dt \\
  & = \alpha^2 e^{-(\tau_k+\tau_l) f_\infty}  \Bigg( \int_0^{\tau_k}  (1 -  (e^{-\alpha \tau_k}+e^{-\alpha \tau_l}) \cosh(\alpha t) + e^{-\alpha(\tau_k + \tau_l)} (1/2+ 1/2 \cosh(2 \alpha t))) \dt  \\
  & \quad\qquad\qquad\qquad +\sinh (\alpha \tau_k) \int_{\tau_k}^{\tau_l}     e^{-\alpha t} -  \frac{1}{2} e^{-\alpha \tau_l}( 1+ e^{-2\alpha t}) \dt \\
  & \quad\qquad\qquad\qquad +\sinh (\alpha \tau_k)  \sinh (\alpha \tau_l) \int_{\tau_l}^{\infty}     e^{-2\alpha t}  \dt \Bigg) \\
  & = \alpha^2 e^{-(\tau_k+\tau_l) f_\infty}  \Bigg( (\tau_k -  (e^{-\alpha \tau_k}+e^{-\alpha \tau_l}) \frac{1}{\alpha} \sinh(\alpha \tau_k) + e^{-\alpha(\tau_k + \tau_l)} (1/2 \tau_k + \frac{1}{4\alpha} \sinh(2 \alpha \tau_k)))  \\
  & \quad\qquad\qquad\qquad +\sinh (\alpha \tau_k)  \left(\frac{-1}{\alpha}   (e^{-\alpha \tau_l}-e^{-\alpha \tau_k}) - \frac{1}{2} e^{-\alpha \tau_l}( \tau_l - \tau_k) + \frac{1}{4 \alpha} (e^{-3\alpha \tau_l}-e^{-\alpha (2\tau_k+\tau_l)})) \right) \\
  & \quad\qquad\qquad\qquad +\sinh (\alpha \tau_k)  \sinh (\alpha \tau_l) \frac{1}{2 \alpha}     e^{-2\alpha \tau_l} \Bigg)
\end{align*}
for the first.

\end{document}